\documentclass{article}
\usepackage{amsmath,amsthm}
\usepackage{amssymb,latexsym}
\usepackage[mathscr]{eucal}
\usepackage{setspace}
\usepackage{graphics}
\usepackage{array}

\DeclareMathSizes{13}{13}{9}{8}

\newcommand{\pa}{\partial}

\newcommand{\rb}{\right]}
\newcommand{\lb}{\left[}
\newcommand{\rel}{\right\}}

\newcommand{\beq}{\begin{equation}}
\newcommand{\eq}{\end{equation}}
\newcommand{\bfv}{{\bf v}}

\newcommand{\bfj}{{\bf J}}
\newcommand{\bfb}{{\bf B}}

\newcommand{\bfe}{{\bf E}}
\newcommand{\bfihat}{\hat {\bf i}}
\newcommand{\na}{\nabla}
\newcommand{\ti}{\times}

\newcommand{\bfve}{\bfv_e}
\newcommand{\bfvi}{\bfv_i}
\newcommand{\vet}{\frac{\pa\bfve}{\pa t}}
\newcommand{\vit}{\frac{\pa\bfvi}{\pa t}}
\newcommand{\pe}{p_e}
\newcommand{\pui}{p_i}
\newcommand{\lc}{\frac{1}{c}}

\usepackage{graphicx}
\graphicspath{{converted_graphics/}}
\begin{document}

\title{Steady Hall Magnetohydrodynamics Near a X-type Magnetic Neutral Line }         
% Enter your title between curly braces
\author{Bhimsen K. Shivamoggi\footnote{\large Permanent Address: University of Central Florida, Orlando, FL 32816-1364}\\
Los Alamos National Laboratory\\
Los Alamos, NM 87545}        % Enter your name between curly braces
\date{}          % Enter your date or \today between curly braces
\maketitle

\large{\bf Abstract}

Hall magnetohydrodynamics  (MHD) properties near a two-dimensional (2D) X-type magnetic neutral line in the steady state are considered via heuristic and rigorous developments. Upon considering the steady-state as the asymptotic limit of the corresponding \textit{time-dependent} problem and using a rigorous development, Hall effects are shown to be able to sustain the hyperbolicity of the magnetic field (and hence a more open X-point configuration) near the neutral line in the steady state. The heuristic development misses this subtle connection of the steady state with the corresponding \textit{time-dependent} problem and predicts only an elongated current-sheet configuration (as in resistive MHD). However, the heuristic development turns out to be useful in providing insight into the lack of dependence of the reconnection rate on the mechanism breaking the frozen-in condition of the magnetic field lines. The latter result can be understood in terms of the ability of the ions and electrons to transport equal amounts of magnetic flux per unit time out of the reconnection region.

\pagebreak

\section{Introduction}      

\indent In resistive magnetohydrodynamics (MHD) the ion inflow is the only means to transport magnetic flux into the reconnection layer (Knoll and Chacon[1]). As the resistivity is decreased, large magnetic pressure gradients which develop upstream of the reconnection layer start inhibiting the ion inflow and the magnetic flux transport into the reconnection layer (and hence the reconnection rate) - the so-called \textit{pressure problem}~(Clark [2]). The Hall effect (Sonnerup [3]) can overcome the \textit{pressure problem} (Dorelli and Birn [4], Knoll and Chacon [5]), thanks to the decoupling of electrons from ions on length scales below the ion skin depth $d_i$. So, if the reconnection layer width is less than $d_i$, the electron inflow can keep on going which transports the magnetic flux into the reconnection layer and hence reduces the flux pile-up. Previous numerical work (Shay et al. [6], Rogers et al. [7], Knoll and Chacon [5]) indicated that the dissipation region in Hall MHD, as $d_i$ increases, changes from an elongated current sheet geometry (Sweet [8]-Parker [9] type) to a more open X-point geometry (Petschek [10] type). However,  recent fully kinetic simulations (Daughton et al. [11], Karimabadi et al. [12]) and EMHD-based treatments (Chacon et al. [13]) have shown that elongated current sheets are also possible. On the other hand, more recent particle-in-cell (PIC) simulations (Shay et al.[14]) show spatial localization of the out-of-plane current to within a few $d_{i}^{,}s$ of the X-line, so this controversy continues. It is therefore in order to shed further light on this issue. In this paper, we consider Hall MHD properties near a two-dimensional (2D) X-type magnetic neutral line in the steady state  via heuristic and rigorous developments and investigate whether or not the Hall effects favor the hyperbolicity of the magnetic field near the neutral line. The heuristic development turns out to be useful in providing insight into the lack of dependence of the reconnection rate on the mechanism breaking the frozen-in condition of the magnetic field lines.

\section{Governing Equations for Hall MHD}
Consider an incompressible, two-fluid, quasi-neutral plasma. The governing equations for this plasma dynamics are (in usual notation) - 

\beq
nm_e \lb \vet + (\bfve\cdot\na)\bfve\rb = -\na\pe-ne (\bfe+\lc\bfve\ti\bfb) + ne\eta \bfj
\eq

\beq
nm_i \lb \vit + (\bfvi\cdot\na)\bfvi\rb = -\na\pui+ne(\bfe+\lc\bfvi\ti\bfb)-ne\eta\bfj
\eq

\begin{align}
\na\cdot\bfve &=0\\
\na\cdot\bfvi &=0\\
\na\cdot\bfb &=0\\
\na\ti\bfb &=\lc\bfj\\
\na\ti\bfe &=-\lc\frac{\pa\bfb}{\pa t}
\end{align}

\noindent
where,
\beq
\bfj\equiv ne(\bfvi - \bfve).
\eq

Neglecting electron inertia ($m_e \rightarrow 0$), equations (1) and (2) can be combined to give an ion equation of motion - 

\beq
nm_i \lb \vit +(\bfvi\cdot\na)\bfvi\rb = -\na(p_i + p_e) +\lc \bfj\ti\bfb
\eq

\noindent
and a generalized Ohm's law - 

\beq
\bfe + \lc \bfv_{i}\ti\bfb = \eta\bfj +\frac{1}{nec}\bfj\ti\bfb.
\eq

Non-dimensionalize distance with respect to a typical length scale $a$, magnetic field with respect to a typical magnetic field strength $B_0$, time with respect to the reference Alfv\'en time $\tau_A \equiv a/V_{A_0}$ where $V_{A_0} \equiv B_0/\sqrt{m_{i}n}$, and introduce the magnetic and velocity stream functions according to

\beq
\left.
\begin{array}{l}
\bfb = \na\psi \ti\bfihat_z + b\bfihat_z\\
\bfvi = \na\phi \ti\bfihat_z + w\bfihat_z
\end{array}\rel
\eq

\noindent
and assume the physical quantities of interest have no variation along the $z$-direction. The Hall magnetic field $b$ is believed to be produced by the dragging of the in-plane magnetic field in the out-of-plane direction by the electron near the X-type magnetic neutral line ([6], [7]). Equations (9) and (10), then yield

\beq
\frac{\pa\psi}{\pa t} + [\psi,\phi] + \sigma [b,\psi] = \hat{\eta} \na^2\psi
\eq

\beq
\frac{\pa b}{\pa t} + [b,\phi] + \sigma [\psi,\na^2\psi] +[\psi, w] = \hat{\eta}\na^2 b
\eq

\beq
\frac{\pa}{\pa t} (\na^2\phi) +[\na^2\phi,\phi] = [\na^2\psi,\psi]
\eq

\beq
\frac{\pa w}{\pa t} + [w,\phi] = [b,\psi]
\eq

\noindent
where,

\beq
\left.
\begin{array}{r}
[A,B] \equiv \na A \ti \na B\cdot\bfihat_z\\
\displaystyle{\sigma \equiv \frac{d_i}{a}, ~\hat{\eta}\equiv\frac{\eta c^2 \tau_A}{a^2}}.\notag
\end{array}
\rel
\eq

\vspace{0.5in}

\noindent
\section{A Heuristic Analysis}

It is instructive to do a heuristic analysis to develop an estimate on the geometry of the dissipation region prior to a more rigorous formulation. Let the dissipation region have a length $L$ in the outflow $x$-direction and a width $\delta$ in the inflow $y$-direction.

We then have from equation (5),
\beq
\frac{B_x}{L}\sim \frac{B_y}{\delta}.
\eq

We have from the $z$-component of equation (10), 

\beq
\eta J_z \sim \frac{1}{nec}J_x B_y
\eq

\noindent and on using equation (6), (17) becomes

\beq
\eta c \frac{B_x}{\delta} \sim \frac{1}{nec} c \frac{B_z}{\delta} B_y\notag
\eq

\noindent or

\beq
\eta B_x \sim \frac{1}{nec} B_y B_z.
\eq

\indent Next, the $z$-component of the curl of equation (10) gives

\beq
\eta c\nabla^2B_z = \frac{1}{nec} (\textbf{B}\cdot\nabla)J_z
\eq
from which,

\beq
\eta c \frac{B_z}{\delta^2}\sim\frac{1}{nec}\frac{B_x}{L}c \frac{B_x}{\delta}
\eq

\noindent and on using (16), (20) becomes

\beq
\eta  \frac{B_z}{\delta^2}\sim\frac{1}{nec}\frac{B_y}{\delta}\frac{B_x}{\delta}\notag
\eq

\noindent or

\beq
\eta B_z \sim \frac{1}{nec}B_x B_y.
\eq

\indent(18) and (21) give 
\beq
B_z \sim B_x.
\eq

\indent Using (22), (18) gives 

\beq
B_y \sim nec\eta.
\eq

\indent Using (16), (23) leads to 
\beq
\frac{\delta}{L} B_x \sim nec\eta.
\eq

On rearranging, (24) gives
\beq
\frac{\delta}{L}\sim \frac{1}{\sigma S}
\eq

\noindent where,

\beq
S\equiv \frac{V_{A_i}a}{\tilde{\eta}},~ \tilde{\eta}\equiv \eta c^2,~ V_{A_i} \equiv \frac{B_x}{\sqrt{nm_i}}\notag .
\eq
(25) was also given by Chacon et al. [13] by using a more rigorous formulation.

\indent Noting that in the Hall resistive regime

\beq
\sigma > 1, ~S >> 1\notag
\eq

\noindent(25) implies that the diffusion region in the Hall resistive region may be expected to be elongated. Rigorous formulation in Section 4, on the other hand, shows that this result represents only part of the story because the heuristic development fails to recognize the steady state as the asymptotic limit of the corresponding \textit{time-dependent} problem (Shivamoggi [15]). The latter aspect makes the actual story more complicated than what is conveyed by the heuristic development. 

\indent It is interesting to note, however,  that the above heuristic analysis sheds some light on the conjecture (Mandt et.al. [16], Shay and Drake [17]) that the reconnection rate in Hall MHD is primarily controlled by ions (which are decoupled from the electrons) and is independent of the mechanism that breaks the frozen-in condition of the magnetic field lines (resistivity or electron inertia).

\indent For the Hall-resistive case, the reconnection rate is give by 

\beq
E \sim \eta J_z
\eq

\noindent and on using equation (6) and (23), (26) becomes 

\beq
E \sim \frac{B_y}
{nec} c \frac{B_x}{\delta}.
\eq

Using (16), (27) becomes
\beq
E \sim \frac{B^2_x}{neL}.
\eq

\indent On the other hand, if we consider the electron inertia to constitute the mechanism that breaks the frozen-in condition of the magnetic field lines, the Ohm's law now takes the form (Coppi et al. [18])

\beq
\textbf{E} + \frac {1}{c} \textbf{v}_e \times \textbf{B} = \frac{1}{\omega^2_{p_e}}\frac{d\textbf{J}}{dt}.
\eq

\indent In the electron-inertia case, the reconnection rate is therefore given by

\beq
E \sim \frac {1}{\omega_{p^2_e}}\frac{dJ_z}{dt}.
\eq

\indent Using equation (6), (30) may be rewritten as

\beq
E \sim \frac {1}{\omega_{p^2_e}}\frac{1}{\tau_{A_e}}\frac{B_x}\delta.
\eq

where, 

\beq
\tau_{A_e} \sim \frac {L}{V_{A_e}},~ V_{A_e} \equiv \frac{B_x}{\sqrt{nm_e}}\notag.
\eq

Taking $\delta \sim d_e$, (31) becomes

\beq
E \sim \frac{B_x{^2}}{n e L}
\eq

\noindent which is the same as the one, namely, (28), for the Hall resistive case! This appears to support the conjecture (Mandt et al. [16], Shay and Drake [17]) that the reconnection rate is independent of the mechanism that breaks the frozen-in condition of the magnetic field lines. A similar conclusion was reached by Chacon et al [13] who considered the electron hyperresistivity to constitute another mechanism that breaks the frozen-in condition of the magnetic field lines.

\indent It is of interest to note that (28) and (32) may be rewritten as 
\beq
E \sim \frac{B^2_x}{neL} \sim B_x  \left(\frac{V_{A_i}}{c}\right)\left(\frac {d_i}{L}\right) \sim B_x \left( \frac{V_{A_e}}{c}\right)\left(\frac{d_e}{L}\right).
\eq

\noindent (33) shows that the lack of dependence of the reconnection rate on the mechanism breaking the frozen-in condition of the magnetic field lines can be understood in terms of the ability of the ions and electrons to transport equal amounts of magnetic flux per unit time out of the reconnection region. The insensitivity of the reconnection rate, according to (33), on the particle mass has been confirmed by the recent partial-in-cell simulations ([14]). We are assuming here, as confirmed by the recent numerical simulations (Drake et al. [19]) that, in the electron-inertia case, the outflow velocity of the electrons from the dissipation region is given by the Alfv\'{e}n speed based on the upstream magnetic field $B_x$ with the width of the dissipation region scaling with $d_e$.

\noindent
\section {Steady-state Properties Near an X-type Neutral Line}

Consider Hall MHD properties near a 2D $X$-type magnetic neutral line in the steady state. Equations (12)-(15) now become

\beq
- c E + \psi_{10} \phi_{01} - \psi_{01} \phi_{10} + \sigma(b_{10} \psi_{01} - b_{01} \psi_{10}) = \hat{\eta} (\psi_{20} + \psi_{02})\\
\eq

\beq
b_{10} \phi_{01} - b_{01} \phi_{10} + \sigma [\psi_{10}(\psi_{21} +\psi_{03}) - \psi_{01} (\psi_{30} + \psi_{12})] + \psi_{10} w_{01} - \psi_{01}w_{10} = \hat{\eta} (b_{20} + b_{02})\\
\eq

\beq
\phi_{10}(\phi_{21} + \phi_{03}) - \phi_{01} (\phi_{30} + \phi_{12}) - [\psi_{10}(\psi_{21} + \psi_{03}) + \psi_{01} (\psi_{30} + \psi_{12})]= -\nu(\phi_{40} + 2 \phi_{22} + \phi_{04})\\
\eq

\beq
w_{10}\phi_{01} - w_{01} \phi_{10} + \psi_{10}b_{01} - \psi_{01}b_{10} = \nu (w_{20} +w_{02})
\eq

\noindent where,

$$
F_{mn}\equiv \frac{\partial^{m+n}F}{\partial x^{m}\partial y^{n}},~E\equiv-~ \frac{1}{c}~\frac{\partial\psi}{\partial t}\notag
$$

\noindent and we have now included in equations (14) and (15) viscous effects in the plasma which become important near the magnetic neutral line (Tsuda and Ugai [20]); $\nu$ is the viscosity coefficient.

\indent Following Cowley [21] and Shivamoggi [22], let us expand the velocity and magnetic fields in a Taylor series about the neutral line taken to be at $x = 0, y = 0$. Equations (34)-(37) may then be used to derive relationships between the coefficents of the series. The latter are simply the partial derivatives of the velocity and magnetic fields at the neutral line. Motivated by the symmetry properties of equations (34) - (37) in the ideal limit, we may consider $\psi$ and $w$ to be even functions of both $x$ and $y$, and $\phi$ and $b$ to be odd functions of both $x$ and $y$ (this also enables the out-of-plane magnetic field $b$ to exhibit the \textit{quadrupolar} structure (Terasawa [23]) characteristic of Hall MHD - this has also been confirmed by laboratory experiments (Ren et al. [24]). Thus, we write 

\beq
\psi = \sum_m \sum_n\Psi_{2m,2n} \frac{x^{2m}y^{2n}}{(2m)! (2n)!}\notag
\eq

\beq
\phi = \sum_m \sum_n \Phi_{2m+1,2n+1} \frac{x^{2m+1} y^{2n+1}}{(2m+1)! (2n+1)!}\notag
\eq

\beq
b = \sum_m \sum_n B_{2m+1, 2n+1} \frac{x^{2m+1} y^{2n+1}}{(2m+1)! (2n+1)!}\notag
\eq

\beq
w = \sum_m \sum_n W_{2m, 2n} \frac{x^{2m} y^{2n}}{(2m)! (2n)!}.
\eq

\noindent(38) reflects the fact that the origin in the $x, y-plane$ is both the X-type neutral point and a stagnation point of the flow.

Using (38), equations (34) and (37) give, on evaluation at the origin,

\beq
\hat{\eta} (\Psi_{20} + \Psi_{02}) = -c E\\
\eq
\beq
\nu (W_{20} + W_{02}) = 0\\
\eq

 Let us differentiate equations (34) and (37) with respect to  $x$ and $y$ separately, and differentiate the resulting four equations, respectively, with respect to $x, y$ and $x,y$. We then obtain

\begin{equation}
\psi_{30} \phi_{01} + 2 \psi_{20} \phi_{11} + \psi_{10} \phi_{21} - \psi_{21} \phi_{10} - 2\psi_{11} \phi_{20} - \psi_{01} \phi_{30}\notag
\end{equation}
\begin{equation}
+ \sigma (b_{30} \psi_{01} + 2b_{20} \psi_{11} + b_{10} \psi_{21} - b_{21} \psi_{10}- 2 b_{11} \psi_{20} - b_{01} \psi_{30})\notag
\end{equation}
\begin{equation} = \hat{\eta} (\psi_{40} + \psi_{22})
\end{equation}
\begin{equation}
\psi_{12} \phi_{01} + 2 \psi_{11} \phi_{02} + \psi_{10} \phi_{03} - \psi_{03} \phi_{10} - 2\psi_{02} \phi_{11} - \psi_{01} \phi_{12}\notag
\end{equation}
\begin{equation}
+ \sigma (b_{12} \psi_{01} + 2b_{11} \psi_{02} + b_{10} \psi_{03} - b_{03} \psi_{10}- 2b_{02} \psi_{11} - b_{01} \psi_{12})\notag
\end{equation}
\beq = \hat{\eta} (\psi_{22}+\psi_{04})
\eq
\begin{equation}
w_{30} \phi_{01} + 2w_{20} \phi_{11} + w_{10} \phi_{21} - w_{21} \phi_{10}- 2w_{11} \phi_{20} - w_{01} \phi_{30}\notag
\end{equation}
\beq
+ \psi_{30} b_{01} + 2\psi_{02} b_{11}+ \psi_{10} b_{21} - \psi_{21} b_{10} - 2\psi_{11} b_{20} - \psi_{01} b_{30}\notag
\end{equation}
\beq
= \nu(w_{40} + w_{22})\\
\eq
\newpage
\begin{equation}
w_{12} \phi_{01} + 2w_{11}\phi_{02} + w_{10} \phi_{03} - w_{03} \phi_{10} - 2w_{02} \phi_{11} - w_{01} \phi_{12}\notag
\end{equation}
\beq
+ \psi_{12} b_{01} + 2\psi_{11} b_{02} + \psi_{10} b_{03} - \psi_{03} b_{10} - 2\psi_{02} b_{11} - \psi_{01} b_{12}\notag
\eq
\beq
= \nu (w_{22} +w_{04}).
\eq

\vspace{0.10in}

Next, let us differentiate both equations (35) and (36) with respect to $x$ and then both with respect to $y$. We then obtain

\beq
b_{21} \phi_{01} + b_{20}\phi_{02} + b_{10} \phi_{12} - b_{12}\phi_{10}
 - b_{02}\phi_{20} - b_{01}\phi_{21}+ \sigma[\psi_{21}(\psi_{21} + \psi_{03}) + \psi_{20} 
(\psi_{22} + \psi_{04})\\\notag
\eq
\beq
+ \psi_{10} (\psi_{32} + \psi_{14}) - \psi_{12} (\psi_{30} +\psi_{12})- \psi_{02} 
(\psi_{40} + \psi_{22}) - \psi_{01}(\psi_{41} + \psi_{23})]\\\notag
\eq
\beq
+ \psi_{21} w_{01} + \psi_{20} w_{02} + \psi_{10}w_{12} - \psi_{12}w_{10} 
- \psi_{02}w_{20}- \psi_{01} w_{21} = \hat{\eta} (b_{31} + b_{13})
\eq
\beq
\phi_{21} (\phi_{21} + \phi_{03}) + \phi_{20} (\phi_{22} + \phi_{04}) + \phi_{10} 
(\phi_{32} + \phi_{14})-\phi_{12} (\phi_{30} + \phi_{12}) - \phi_{02} 
(\phi_{40} + \phi_{22}) - \phi_{01} (\phi_{41} + \phi_{23})\\\notag
\eq
\beq
= \psi_{21} (\psi_{21} + \psi_{03}) + \psi_{20} (\psi_{22} + \psi_{04}) + \psi_{10} 
(\psi_{32} + \psi_{14})- \psi_{12} (\psi_{30} + \psi_{12}) - \psi_{02}
(\psi_{40} + \psi_{22}) - \psi_{01} (\psi_{41} + \psi_{23})\\\notag
\eq
\beq
-\nu (\phi_{51} + 2\phi_{33} + \phi_{15}).\\
\eq

\vspace{0.10in}
We now use (38) and evaluate equations (41) - (46) at the origin:
\beq
\hat{\eta} (\Psi_{40} + \Psi_{22}) = 2 \Psi_{20}(\Phi_{11} - \sigma B_{11})\\
\eq

\beq
\hat{\eta}(\Psi_{22} + \Psi_{04}) = -2\Psi_{02}(\Phi_{11} - \sigma B_{11})\\
\eq

\beq
\hat{\eta} (B_{31} + B_{13}) = \sigma [\Psi_{20} (\Psi_{22} + \Psi_{04})
-\Psi_{02} (\Psi_{40} + \Psi_{22})]\\\notag
\eq

\beq
+ \Psi_{20}W_{02} - \Psi_{02}W_{20}\\
\eq

\beq
\Psi_{20}(\Psi_{22} + \Psi_{04}) - \Psi_{02} (\Psi_{40} + \Psi_{22})
= \nu (\Phi_{51} + 2 \Phi_{33} + \Phi_{15})\\
\eq

\beq
\nu (W_{40} + W_{22}) = 2 W_{20} \Phi_{11} + 2 \Psi_{02} B_{11}.\\
\eq

\beq
\nu (W_{22} + W_{04}) = -2W_{02} \Phi_{11} - 2 \Psi_{20} B_{11}
\eq

\vspace{0.05in.}
Using equations (47) and (48), equation (50) gives
\beq
4(\Phi_{11} - \sigma B_{11}) \Psi_{20}\Psi_{02} = -\nu\hat{\eta} (\Phi_{51} + 2 \Phi_{33} + \Phi_{15}).\\
\eq

Equation (53) shows that, in the MHD resistive viscous case $(\sigma = 0, ~\hat{\eta} ~ \text{and}~\nu\ \not= 0)$, one has 

\beq
\Psi_{20} \not= 0, \Psi_{02}\not=0\\
\eq
so the magnetic field to lowest order can be hyperbolic (Shivamoggi [22]). On the other hand, in the inviscid or non-resistive case, when Hall effects are included ($B_{11} \not= 0$ - Hall effects materialize only via their signature - the \textit{quadrupolar} out-of-plane magnetic field pattern), on considering the steady state as the asymptotic limit of the corresponding \textit{time-dependent} problem\footnote{\large Indeed, one way to resolve difficulties that arise in insuring uniqueness of solution of stationary \textit{linear} wave problems is to pose a more realistic \textit{unsteady} problem with suitable initial conditions applied at some finite time $t = -t_o,$ say, in the past, and then letting $t_o \Rightarrow \infty$ in the solution (Lighthill [25]).} (Shivamoggi [15]), we have from the latter formulation 
\vspace{0.10in}

\noindent  
\beq
\Phi_{11} -\sigma B_{11} = 0.
\eq
\vspace{0.10in}

\noindent Using (55), equation (53) shows that, even in the inviscid limit $(\nu \Rightarrow 0)$, equation (54), thanks to Hall effects, continues to be valid. So, Hall effects are able to sustain the hyperbolicity of the magnetic field (and hence a more open \textit{X}-point configuration) near the neutral line. Equation (55) implies that the level curves of the out-of-plane magnetic field are also the streamlines of the in-plane ion flow.

\indent Further, using (55), equations (49), (51) and (52) lead to the following compatibility conditions on the Taylor expansion coefficients of the out-of-plane components of the velocity and magnetic fields - 

\beq
\hat{\eta} (B_{31} + B_{13}) = 0\\
\eq

\beq
\Psi_{02} + \sigma W_{02} = 0\\
\eq

\beq
\Psi_{20} + \sigma W_{20} = 0.\\
\eq

\section{Discussion}

\indent In this paper, we have considered Hall MHD properties near a 2D X-type magnetic neutral line in the steady state via heuristic as well as rigorous developments. Upon considering the steady state as the asymptotic limit of the corresponding \textit{time-dependent} problem and using a rigorous development, Hall effects are shown to be able to sustain the hyperbolicity of the magnetic field (and hence a more open X-point configuration) near the neutral line in the steady state. The heuristic development misses this subtle connection of the steady state with the corresponding \textit{time-dependent} problem and predicts only an elongated current-sheet configuration (as in resistive MHD). On the other hand, this development also shows that the electron-hyperresistivity effects are similar to that of ion viscosity $\nu$ and allow for the possibility of a X-point magnetic field configuration (see Appendix) as found also by Chacon et al. [13].
The heuristic development, however, turns out to be useful in providing insight into the lack of dependence of the reconnection rate on the mechanism breaking the frozen-in condition of the magnetic field lines. The latter result can be understood in terms of the ability of the ions and electrons to transport equal amounts of magnetic flux per unit time out of the reconnection region.
\vspace{0.05in}
\section{Acknowledgements}

\noindent I acknowledge with gratitude the stimulating interactions and discussions with Dr. Luis Chacon that led to this work. My thanks are due to Drs. Michael Shay and Michael Johnson for helpful communications and discussions.

\section{Appendix}

On including the effects of ion viscosity and electron hyperresistivity, equations (12)- (15) become

\beq
\frac{\pa\psi}{\pa t} + [\psi,\phi] + \sigma [b,\psi] = \hat{\eta} \na^2\psi + \eta_h \na^4\psi\\
\eq

\beq
\frac{\pa b}{\pa t} + [b,\phi] + \sigma [\psi,\na^2\psi] +[\psi, w] = \hat{\eta} \na^2 b + \eta_h \na^4 b\\
\eq

\beq
\frac{\pa}{\pa t} (\na^2\phi) +[\na^2\phi,\phi]+[\psi,\na^2\psi]= \nu \na^4\phi\\
\eq

\beq
\frac{\pa w}{\pa t} + [w,\phi] +[\psi, b] = \nu \na^2 w\\
\eq

\noindent where $\eta_h$ is the electron hyperresistivity coefficient.

\indent Following through the development outlined in Section 4, we have in place of equations (41),~(42) and (45), respectively,

\begin{equation}
\psi_{30} \phi_{01} + 2 \psi_{20} \phi_{11} + \psi_{10} \phi_{21} - \psi_{21} \phi_{10} - 2\psi_{11} \phi_{20} - \psi_{01} \phi_{30}\notag
\end{equation}

\begin{equation}
+ \sigma (b_{30} \psi_{01} + 2b_{20} \psi_{11} + b_{10} \psi_{21} - b_{21} \psi_{10}- 2 b_{11} \psi_{20} - b_{01} \psi_{30})\notag
\end{equation}

\begin{equation} = \hat{\eta} (\psi_{40} + \psi_{22})+ \eta_h (\psi_{60} + \psi_{24})
\end{equation}

\begin{equation}
\psi_{12} \phi_{01} + 2 \psi_{11} \phi_{02} + \psi_{10} \phi_{03} - \psi_{03} \phi_{10} - 2\psi_{02} \phi_{11} - \psi_{01} \phi_{12}\notag
\eq

\begin{equation}
+ \sigma (b_{12} \psi_{01} + 2b_{11} \psi_{02} + b_{10} \psi_{03} - b_{03} \psi_{10}- 2b_{02} \psi_{11} - b_{01} \psi_{12})\notag
\end{equation}

\beq = \hat{\eta} (\psi_{22}+\psi_{04})+ \eta_h (\psi_{42} + \psi_{06})
\eq

\beq
b_{21} \phi_{01} + b_{20}\phi_{02} + b_{10} \phi_{12} - b_{12}\phi_{10}
 - b_{02}\phi_{20} - b_{01}\phi_{21}+ \sigma[\psi_{21}(\psi_{21} + \psi_{03}) + \psi_{20} 
(\psi_{22} + \psi_{04})\\\notag
\eq

\beq
+ \psi_{10} (\psi_{32} + \psi_{14}) - \psi_{12} (\psi_{30} +\psi_{12})- \psi_{02} 
(\psi_{40} + \psi_{22}) - \psi_{01}(\psi_{41} + \psi_{23})]\\\notag
\eq

\beq
+ \psi_{21} w_{01} + \psi_{20} w_{02} + \psi_{10}w_{12} - \psi_{12}w_{10} 
- \psi_{02}w_{20}- \psi_{01} w_{21} = \hat{\eta} (b_{31} + b_{13})+ \eta_h (b_{51} + b_{15}).
\eq

\vspace{0.15in}

Using (38) and evaluating equations (64) - (66) at the origin,
\vspace{0.10in}
\beq
\hat{\eta} (\Psi_{40} + \Psi_{22})+ \eta_h(\Psi_{60} + \Psi_{24}) = 2 \Psi_{20}(\Phi_{11} - \sigma B_{11})
\eq

\beq
\hat{\eta}(\Psi_{22} + \Psi_{04})+ \eta_h(\Psi_{42} + \Psi_{06}) = -2\Psi_{02}(\Phi_{11} - \sigma B_{11})
\eq

\beq
\hat{\eta} (B_{31} + B_{13})+ \eta_h(B_{51} + B_{15}) = \sigma [\Psi_{20} (\Psi_{22} + \Psi_{04})
-\Psi_{02} (\Psi_{40} + \Psi_{22})] + \Psi_{20} W_{02} - \Psi_{02} W_{20}.
\eq
\vspace{0.15in}

Using equations (67) and (68), equation (50) gives
\vspace{0.05in}
\begin{equation}
4 (\Phi_{11} - \sigma B_{11}) \Psi_{20}\Psi_{02} = - \nu \hat{\eta}(\Phi_{51} + 2 \Phi_{33} + \Phi_{15})- \eta_h (\Psi_{60} + \Psi_{42} + \Psi_{24} + \Psi_{06}).
\end{equation}
\vspace{0.02
in}

It is of interest to note that the right hand side in equation (70) comes from the term - 
\beq
\left(\frac{\partial^2}{\partial x^2}+ \frac{\partial^2}{\partial y^2}\right)\left[\nu\hat{\eta}\left(\frac{\partial^4}{\partial x^3}{\partial y} + \frac{\partial^4}{\partial x \partial y^3}\right) \phi + \eta_h \left(\frac{\partial^4}{\partial x^4} + \frac{\partial^4}{\partial y^4}\right)\psi\right]\notag
\end{equation}
\vspace{0.15in}

\noindent which highlights the similarity between the ion-viscosity and electron hyperresistivity contributions - both allow the possibility of a $X$-point magnetic field configuration.

\end{document}